\documentstyle[referee]{mn}

\newcommand{\reference}{\bibitem}
\def\mnras{MNRAS}
\def\araa{ARAA}
\def\aap{A\&A}
\def\apj{ApJ}
\def\plotone#1{\centering \leavevmode
\epsfxsize=\columnwidth \epsfbox{#1}}

\def\beq{\begin{equation}}
\def\eeq{\end{equation}}
\def\bey{\begin{eqnarray}}
\def\eey{\end{eqnarray}}
\def\kms{{\rm \,km\,s^{-1}}}

\def\kpc{\,{\rm {kpc}}}

\def\chisq{\chi^2}

\def\tp{t_{0}}
\def\tE{t_{\rm E}}
\def\fl{f_{\rm s}}

\def\rEt{\tilde{r}_{\rm E}}

\def\rE{r_{\rm E}}

\def\u0{u_0}

\def\Ax{{\cal A}_x}
\def\Ay{{\cal A}_y}

\def\mI0{m_{I,0}}
\def\Mp{M^\prime}

\input epsf
\begin{document}

\title[Acceleration and Parallax Effects in Gravitational Microlensing]
{Acceleration and Parallax Effects in Gravitational Microlensing}

\author[Smith, Mao \&  Paczy\'nski]
{Martin C. Smith$^1$, Shude Mao$^1$, Bohdan Paczy\'nski$^2$
\thanks{e-mail: (msmith, smao)@jb.man.ac.uk, bp@astro.princeton.edu}
\\
\smallskip
$^{1}$Univ. of Manchester, Jodrell Bank Observatory, Macclesfield,
Cheshire SK11 9DL, UK \\ 
$^{2}$Princeton University Observatory, Princeton, NJ 08544-1001, USA
}
\date{Accepted ........
      Received .......;
      in original form ......}

\pubyear{2002}

\maketitle
\begin{abstract}
To generate the standard microlensing light curve one assumes that the
relative motion of the source, the lens, and the observer is
linear. In reality, the relative
motion is likely to be more complicated due to accelerations of the
observer, the lens and the source. The simplest approximation beyond
the linear-motion assumption is to add a constant acceleration. 
Microlensing light curves due to accelerations can be symmetric
or asymmetric depending on the angle between the acceleration and
the velocity. We show that it is possible that some of the previously
reported shorter marginal parallax events can be reproduced with
constant-acceleration models, while the longer, multi-year parallax
events are ill-fitted by such models.
We find that there is a generic degeneracy inherent in
constant-acceleration microlensing models.
We also find that there is an equivalent degeneracy in parallax
models, which manifests itself in short-duration events.
The importance of this new parallax degeneracy is illustrated with an
example, using one of these marginal parallax events.
Our new analysis suggests that another of these previously
suspected parallax candidate events may be exhibiting some weak
binary-source signatures. If this turns out to be true, 
spectroscopic observations of the source could determine
some parameters in the model and may also constrain or even determine
the lens mass. We also point out that symmetric
light curves with constant accelerations can mimic
blended light curves, producing misleading
Einstein-radius crossing time-scales when fitted by the standard
`blended' microlensing model; this may have some effect on the
estimation of optical depth.
\end{abstract}

\begin{keywords}
gravitational lensing - Galaxy: bulge - Galaxy: centre - binaries: general
\end{keywords}

\section{Introduction}

A decade ago Gould (1992) suggested that the Earth's orbital motion
may affect long duration events of gravitational microlensing,
producing a periodic modulation of the event light curve.  The first
such example of this behaviour was detected by Alcock et al. (1995),
and a number of additional events have been reported by several groups
(e.g., Mao 1999; Smith, Mao \& Wo\'zniak 2002a;
Bennett et al. 2002a; Bond et al. 2001). However, most of these events
were of relatively short duration and showed a strong asymmetry in the  
light curves, rather than a yearly modulation.  Two events were long 
enough to make a yearly modulation clearly noticeable: OGLE-1999-BUL-32 =
MACHO-99-BLG-22 (Mao et al. 2002; Bennett et al. 2002b) and OGLE-1999-BUL-19
(Smith et al. 2002b).

While the Earth's motion may generate a photometric parallax effect, a
similar anomaly may be due to binary motion of the source (sometimes
referred to as `xallarap'; Griest \& Hu 1992; Han \& Gould 1997;
Paczy\'nski 1997) and/or the lens.  In fact such a binary-source
modulation was detected in the microlensing event MACHO-SMC-1 (Alcock
et al. 1997; Palanque-Delabrouille et al. 1998; Udalski et al. 1997) and
MACHO 96-LMC-2 (Alcock et al. 2001). While in these two 
cases the binary source period was much shorter than the event time scale,
an alternative situation may also be encountered.  In fact, Smith et
al. (2002b) pointed out that even in the case of a very long event
exhibiting an annual modulation of its light curve, a
model based on a binary source moving with a one year orbital period
and suitable inclination and orientation of the orbit may generate a
light curve identical to that expected from the parallax effect.
While such a coincidence is very unlikely, it is also verifiable in
that a binary source should display periodic variations in its radial
velocities.

As most parallax microlensing events are not very long it is likely that
the effects they show may be well described with a constant acceleration of
the Earth, the lens and/or the source, with no need to invoke a specific
orbit. The aim of this paper is to investigate this possibility and to
present several consequences of the proposed approach. In Section 2 we
describe a simple model that incorporates a constant-acceleration
term into the microlensing formalism. We then test the efficacy of
this model in Section 3 by fitting a number of microlensing events
that have been previously identified as exhibiting variations from
the standard constant-velocity model of microlensing. The events that
we have selected were all identified as potential parallax
microlensing events, and in this section we aim to test whether this 
constant-acceleration model is able to successfully reproduce this
behaviour. In this section we also mention that there is a generic
degeneracy with the constant-acceleration model, and that there may
be an equivalent degeneracy for parallax models.
In Section 4 we summarise our work and discuss some of the
implications of our findings. In particular, we note that the effect of
constant-acceleration perturbations is likely to be of little
importance for binary lenses, and such perturbations are therefore
most-likely to be due to accelerations of the observer and/or the source.

\section{A simple Model}

In the simplest case of a microlensing event, the light curve is calculated
assuming that the relative motion of the source, the lens, and the observer
is linear and constant 
(Paczy\'nski 1986).  The next step in allowing for a more 
complicated relative motion is to expand it in a power series, and to
retain the first two terms, i.e. the velocity, $ {\bf v} $, and the
acceleration, $ {\bf a }$. For the purposes of this paper, we define
the velocity, ${\bf v}$, to be the velocity of the lens relative to the
observer-source line-of-sight. This acceleration is therefore a
relative acceleration, but since the following analysis is likely to
be valid only for lenses in inertial motion (see Section \ref{disc}),
this acceleration can be thought of as the acceleration of the
observer and/or the source.
An acceleration related velocity vector may 
be defined as $ {\bf v_a } = {\bf a } \tE $, where $ \tE $ is the event
time scale: $ \tE = \rE / v $, $ \rE $ is the Einstein radius and
$v$ is the velocity at the minimum separation (peak magnification).
We expect that any `anomaly' in the light curve will depend on two 
dimensionless numbers: the ratio of the two velocities: $ {\cal A} = v_a/v $,
and the angle between them $\theta$.
Notice
${\cal A}$ is simply the physical acceleration in units of $\rE/\tE^2$. 

In general, an acceleration will produce asymmetry in the microlensing
light curve, and this is the effect which is used to select parallax
events.  However, if the acceleration is mostly in the direction
perpendicular to the velocity vector $ {\bf v } $ then there will be
no asymmetry, just a change in the shape that may be difficult to
notice.

Fig. \ref{fig:model} shows four example light curves with different
accelerations. All four cases have a minimum impact parameter of
$u_0=0.1$, and the source is assumed to move along the $x$-axis
in the absence of acceleration. The dimensionless acceleration parameters
along the $x$ and $y$ directions, $\Ax$ and $\Ay$, are indicated in each panel.
The top left panel shows a light curve in which the acceleration is $45^\circ$
from the $x$-axis while the bottom left panel shows an example
where the acceleration is in the same direction as the source motion.
As expected, both light curves show clear asymmetries. 
The two right panels show light curves in which the acceleration
is perpendicular to the $x$-axis. The light curves are
symmetric in such cases. The effect of acceleration may not
be noticed, and a standard fit forced upon the event may generate incorrect
values of the impact parameter and the event time scale; we return
to this issue briefly in the discussion.  A thorough analysis
of all such possibilities in current experiments
is beyond the scope of this paper.  Instead, we
concentrate on the readily available candidate parallax events, in
order to check how many of them can be well fitted by adding an
acceleration vector to the standard model.

\section{Analysis of Candidate Parallax Microlensing Events Using
A Constant Acceleration Model}

In this section we implement the above constant-acceleration
microlensing model and test its efficacy by fitting a number of
microlensing events. By analysing previously identified candidate
parallax microlensing events, we aim to investigate whether this
constant-acceleration model is able to suitably reproduce the
observed (i.e. seemingly parallactic) behaviour.

\subsection{Constant Acceleration Model}
\label{model}

We shall first outline the standard constant-velocity microlensing
model. For the standard microlensing light curve, the
source position (in units of Einstein radius) is given by
\beq \label{eq:linear}
x = {t - \tp \over \tE } \equiv x_0(t), ~~~ y = u_0,
\eeq
where $\tp$ is the time of maximum magnification and $u_0$ is
the corresponding (minimum) impact parameter. The magnification
is given by (Paczy\'nski 1986)
\beq \label{eq:mag}
A(t) = {u^2+2 \over u \sqrt{u^2+4}},~~~ u=\sqrt{x^2+y^2}.
\eeq
With acceleration, the $(x, y)$ coordinates must
be modified to
\beq \label{eq:pos}
x = x_0(t) + {1 \over 2} ~{\cal A} \cos\theta ~ x^2_0(t), ~~~
y=u_0 + {1 \over 2} ~ {\cal A} \sin\theta ~ x^2_0(t),
\eeq
where $x_0(t)$ is defined in eq. (\ref{eq:linear}),
${\cal A}$ is the physical acceleration in units of $\rE/\tE^2$, and
the magnification is again calculated using eq. (\ref{eq:mag}).
Notice that, since the transverse velocity changes,
$\tE$ should be understood as the Einstein radius crossing
time corresponding to the transverse velocity $v$ at time $\tp$.
Also, in general, $u_0$ may no longer rigorously be
the minimum impact parameter, but may still provide
a useful approximation, particularly if ${\cal A}$ is small.

To fit a microlensing light curve with the standard model, one
needs a minimum of four parameters, $u_0$, $\tp$, $\tE$, and the
baseline flux (or magnitude), $f_0$. One also often
requires an extra blending parameter, which
accounts for the additional light contributed by the
lens and/or other stars in the crowded field. To do this, we
use a parameter, $\fl$, defined as the fraction of light
contributed by the lensed source at the baseline. To incorporate the
constant acceleration, we need two extra parameters, ${\cal A}$ and
the angle $\theta$ (or equivalently, $\Ax$ and $\Ay$).

We look for any degeneracy in the parameter space for this
constant-acceleration model by fitting each event 1000 times, with the
initial parameter guesses selected from wide range of randomly 
chosen values. We find that two sets of parameters exist for each
$\chisq$ minimum. The parameters are identical except for the
time-scale, $\tE$, and the acceleration parameters, ${\bf {\cal A}}$
and $\theta$. However, the magnitude of the physical acceleration in
units of $\rE/{\rm day}^2$ (i.e., $\left| {\bf a}/\rE \right| ={\bf
{\cal A}}/\tE^2$, as defined in \S2) is the same for both fits,
although the direction of this vector (i.e., $\theta$) differs.
In Appendix A1, we show that this degeneracy can be understood 
analytically. An example of this degeneracy is illustrated in the inset
of Fig. \ref{lc:sc6} for the microlensing event sc6\_2563;
even though the two trajectories look very different they both result
in identical light curves. The full analysis of the event sc6\_2563 is
described in Section \ref{sc6}.

It should be noted that one may expect this degeneracy to manifest
itself in parallax models. In particular, this could be
important for short-duration events that exhibit weak parallax
signatures, since in this case the Earth's acceleration can be
approximated to be constant (i.e. the regime in which this degeneracy
is expected to occur). By testing this hypothesis on the event
sc6\_2563, we are able to verify the existence of this degeneracy.
However, it seems that this may only become apparent for weak
parallax candidates, since a more convincing parallax candidate
(sc33\_4505; Section \ref{sc33}) shows no such degeneracy (see Appendix 
A). A full discussion of this degeneracy, including the application to events
sc6\_2563 and sc33\_4505, can be found in Appendix A2.

\subsection{Analysis of candidate parallax microlensing events}

We proceed by analysing a selection of events that have previously
been identified as candidate parallax microlensing events. 
The parallax model used here requires 7 parameters: the 5 parameters
from the standard model plus two additional parameters to describe
the lens trajectory in the ecliptic plane, namely the Einstein radius
projected onto the observer plane, $\rEt$, and an angle $\psi$ in the
ecliptic plane. $\psi$ is defined  as the angle between the
heliocentric ecliptic $x$-axis and the normal to the trajectory (This
geometry is illustrated in Fig.~5 of Soszy\'nski et al. 2001).
These candidate parallax events have
been selected to provide a broad range of parallactic signatures,
ranging from marginal cases to those displaying more prominent effects.

\subsubsection{sc6\_2563}
\label{sc6}

\begin{table*}
\begin{center}
\caption{The best standard model (first row), the best parallax
models (second and third rows), and the best constant-acceleration
model (fourth and fifth rows) for the OGLE-II event sc6\_2563. The
final column shows the $\chisq$ and the number of degrees of freedom
(dof) for each model. The parameters are explained in \S2 \&
\S3. $\theta$ is given in units of radians. There are two sets of
constant-acceleration parameters since each $\chisq$ minimum has two
degenerate fits with differing values for $\tE$, ${\cal A}$ and
$\theta$. There are also two sets of parallax parameters, owing to the
parallax degeneracy that is analogous to this constant-acceleration
degeneracy. Both degeneracies are discussed in Section \ref{model}
and Appendix A.}
\label{tb:sc6}
\tiny
{
\tabcolsep=0.4\tabcolsep
\begin{tabular}{ccccccccccc}
Model & $t_0$ & $\tE$ (day) & $\u0$ & $f_0$
& $\fl$ & $\psi$ & $\rEt$ (au) & ${\cal A}$ & $\theta$ &
$\chisq$/dof \\
\hline
S &
$   1251.971\pm 0.054 $ &
$     89.2\pm 2.2 $ &
$      0.0758^{+ 0.0026}_{-0.0025} $ &
$    169.43\pm 0.41 $ &
$      0.802^{+ 0.026}_{-0.025} $ &
--- & --- &
--- & --- &
408.7/216
\\ \\
P & 
$   1257.4^{+  4.1    }_{ -1.1    } $ &
$     71.5^{+  1.8    }_{ -1.3    } $ &
$      0.1289\pm 0.0057 $ &
$    169.98^{+ 0.41    }_{-0.42    } $ &
$      1.000^{+    0    }_{-0.022} $ &
$      0.134^{+ 0.089}_{-0.026} $ &
$      4.021\pm 0.086 $ &
--- & --- &
347.4/214
\\ \\
$\rm{P}^\prime$ & 
$   1247.6^{+ 1.1    }_{ -3.7    } $ &
$     73.4^{+ 1.2    }_{ -0.7    } $ &
$      0.368^{+ 0.074}_{- 0.057} $ &
$    169.91 \pm 0.41 $ &
$      1.000^{+    0    }_{-0.020} $ &
$      0.129^{+ 0.091}_{-0.025} $ &
$      3.8^{+ 1.9}_{- 0.9} $ &
--- & --- &
348.2/214
\\ \\
A &
$   1251.774\pm 0.057 $ &
$     71.78^{+2.39}_{-0.52} $ &
$      0.09253^{+0.00063}_{-0.00353} $ &
$    170.32\pm 0.41 $ &
$      1.000^{+0}_{-0.040} $ &
--- & --- &
$      0.496^{+0.065}_{-0.069} $ &
$     4.233^{+0.053}_{-0.050} $ &
332.6/214
\\ \\
$\rm{A}^\prime$ &
$   1251.774\pm 0.057 $ &
$     74.88^{+  2.55    }_{-0.45    } $ &
$      0.09253^{+ 0.00063}_{-0.00353} $ &
$    170.32\pm 0.41 $ &
$      1.000^{+0}_{-0.041} $ &
--- & --- &
$     0.539^{+ 0.082}_{-0.079} $ &
$      2.072^{+ 0.049}_{-0.055} $ &
332.6/214
\end{tabular}
}
\end{center}
\end{table*}

This event was detected by the OGLE collaboration (Udalski et
al. 2000). Its asymmetric nature was
identified during a parallax search of the OGLE-II database (Wo\'zniak
et al. 2001), reported in Smith et al. (2002a). It was
classified as a `marginal' candidate, since the deviations from the
standard model are not particularly pronounced and the improvement in
$\chi^2$ is only slight. The best-fit parameters for the standard and
parallax models are given in Table \ref{tb:sc6}, and the corresponding
light curves are plotted in Fig. \ref{lc:sc6}. The inset of this
figure shows the two degenerate trajectories for the
constant-acceleration model; as was discussed above, each $\chisq$
minimum has two degenerate fits. Both of these trajectories produce
identical light curves, even though the parameter values for $\tE$,
${\cal A}$ and $\theta$ differ. This table also includes an additional
parallax fit, which has a $\chisq$ difference of 0.8 compared to the
best-fit parallax model. This is a manifestation of the parallax
degeneracy that is analogous to this constant-acceleration degeneracy.
See Section \ref{model} and Appendix A for a discussion of these
degeneracies.

It would be expected that a constant-acceleration fit may be suitable
for this event since the duration is not very long, and the asymmetry
shows no signs of modulation. Indeed, when this event is fit with the
above constant-acceleration model the best-fit $\chi^2$ value improves
upon the best-fit parallax value from 347.4 to 332.6. If the error
bars are rescaled so that the best-fit $\chisq$ per degree of freedom
is renormalised to unity, then the difference in $\chisq$ between
these two fits is 9.5 for no additional free parameters, i.e. a
significant 3-$\sigma$ improvement.
The best-fit parameters
and corresponding light curve for the acceleration model are given in
Table \ref{tb:sc6} and Fig. \ref{lc:sc6}, respectively. From this
figure it is clear that the parallax and constant-acceleration models
produce similar fits. However, both models seem to over-predict
the flux around $1000 < t < 1150$ days, implying that there may be
some non-constant contribution to the acceleration that is not
parallactic in nature. Notice that the standard microlensing model
predicts a blending parameter $\fl \approx 0.8$ while other
models predict $\fl \approx 1$. Correspondingly, the Einstein radius
crossing time is about 25\% larger for the standard model than
for the other models; this has implications for the
optical depth estimate in the experiments (see Section \ref{disc}).

\subsubsection{OGLE-1999-CAR-1}

This event was detected in real-time by the OGLE Early Warning
System\footnote{http://www.astrouw.edu.pl/\~{}ogle/ogle3/ews/ews.html}
(Udalski et al. 1994) toward the Carina spiral arm. It has been found
to exhibit systematic deviations from the standard model (Mao
1999). The duration for this event is well over 100 days, and it was
concluded that these deviations are due to the parallax effect.

As with sc6\_2563, the deviations from the standard model for
OGLE-1999-CAR-1 show no signs of modulation, and so one would expect
that this event may be suitably approximated by a constant
acceleration model. This appears to be correct, since the best-fit
constant-acceleration model is able to prove a very similar fit (with
a slightly worse $\chisq$ value of 619.4 compared with
the parallax best-fit value of 614.8). The
best-fit parameters for all models are given in Table \ref{tb:car},
and the corresponding light curves are shown in Fig. \ref{lc:car}. It
is interesting to note that both of the degenerate acceleration fits
have values of ${\cal A} = v_a/v$ greater than 1 (${\cal A} = 1.4$,
and ${\cal A} = 2.4$). This implies that $v_a$, the change in velocity
due to this constant acceleration in a time $\tE$, is greater than the
$v$, the velocity at the point of closest approach.

Also of note is the difference in $\tE$ among the three models. The
values range from $\tE = 126$ days for the parallax model, 
to as much as $\tE = 190$ days for the constant-acceleration
model. These differences in the time-scale are closely connected
with their differences of the blending parameter, due to 
potential degeneracy between these two parameters (Wo\'zniak \&
Paczy\'nski 1997).

\begin{table*}
\begin{center}
\caption{The best standard model (first row), the best parallax
model (second row), and the best constant-acceleration model (third
and fourth rows) for OGLE-1999-CAR-1. The parameters are explained in \S2 \&
\S3. $\theta$ is given in units of radians. There are two sets of
constant-acceleration parameters since each $\chisq$ minimum has two
degenerate fits with differing values for $\tE$, ${\cal A}$ and
$\theta$ (see Section \ref{model} and Appendix A1).}
\label{tb:car}
\tiny
{
\tabcolsep=0.4\tabcolsep
\begin{tabular}{ccccccccccc}
Model & $t_0$ & $\tE$ (day) & $\u0$ & $f_0$
& $\fl$ & $\psi$ & $\rEt$ (au) & ${\cal A}$ & $\theta$ &
$\chisq$/dof \\
\hline
S &
$   1284.05\pm 0.13 $ &
$    130.7^{+ 6.9}_{-6.4} $ &
$      0.207^{+ 0.014}_{-0.013} $ &
$     17.9681\pm 0.0021 $ &
$      0.525^{+ 0.043}_{-0.040} $ &
--- & --- &
--- & --- &
767.1/722
\\ \\
P & 
$   1285.4^{+ 1.3}_{-1.7} $ &
$    126^{+15}_{-10} $ &
$      0.281^{+ 0.019}_{-0.018} $ &
$     17.9677\pm 0.0021 $ &
$      0.48^{+ 0.12}_{-0.11} $ &
$      1.22^{+ 0.33}_{-0.21} $ &
$      7.8^{+ 1.7}_{-1.3} $ &
--- & --- &
614.8/720
\\ \\
A &
$   1282.46\pm 0.20 $ &
$    144^{+29}_{-17} $ &
$      0.162^{+ 0.030}_{-0.033} $ &
$     17.9654\pm 0.0020 $ &
$      0.391^{+ 0.087}_{-0.091} $ &
--- & --- &
$     1.37^{+ 0.37}_{-0.61} $ &
$      4.449^{+ 0.035}_{-0.044} $ &
619.4/720
\\ \\
$\rm{A}^\prime$ &
$   1282.46\pm 0.20 $ &
$    190^{+52}_{-31} $ &
$      0.162^{+ 0.030}_{-0.033} $ &
$     17.9654\pm 0.0020 $ &
$      0.391^{+ 0.087}_{-0.091} $ &
--- & --- &
$      2.38^{+ 1.50}_{-0.81} $ &
$      1.922^{+ 0.041}_{-0.033} $ &
619.4/720
\end{tabular}
}
\end{center}
\end{table*}

\subsubsection{sc33\_4505}
\label{sc33}

This OGLE-II event was also found during the search of Smith et
al. (2002a), but was classified as a `convincing' parallax candidate.
Unlike sc6\_2563, the duration of this event was much longer
($\tE\sim200$ days, cf. $\tE\sim100$ days for sc6\_2563), and the data
show clear signs of deviation from the standard model in both the
rising and declining branch of the light curve. Therefore, if this
asymmetry is indeed due to the parallax effect then one would suspect
that a constant-acceleration model would be unable to reproduce this
behaviour (since the acceleration of the Earth cannot be considered
constant during this time span). On the other hand, if this asymmetry
is due to other causes, the constant-acceleration model may provide a
better fit.

The best-fit parameters for the standard and parallax models are given
in Table \ref{tb:sc33}, and the corresponding light curves are shown
in Fig. \ref{lc:sc33}. A problem arises when this event is fit with
the constant-acceleration model, in that two minima are identified,
with $\chi^2=187.3$ and $\chi^2=227.3$,
respectively. Due to the degeneracy discussed in \S3 and Appendix A1,
there are two degenerate fits corresponding to each of these two
$\chi^2$ values. The two degenerate fits that correspond to $\chi^2=187.3$
have extremely large time-scales
($\tE=1812.8\,{\rm day}$ and $\tE=4880.5\,{\rm day}$),
and the lensed star contributes only
2.3 per cent of the total baseline light (i.e., $\fl=0.023$). 
These values do not seem physical to us, and therefore the fits
corresponding to this $\chisq$ minimum are disregarded.
The best-fit parameters for the two degenerate fits
corresponding to $\chi^2=227.3$ are more `feasible'; they are given in
Table \ref{tb:sc33} and the corresponding light curve is shown in
Fig. \ref{lc:sc33}.
 
However, the discarded `unfeasible' constant-acceleration models
have an overall best $\chisq$ value (187.3). Since this $\chisq$ value is
better than the best-fit parallax model, it highlights that there
could be a deficiency in the parallax fit. This deficiency in the
parallax fit can be seen in the residual plot in Fig. \ref{lc:sc33};
systematic residuals from the parallax model can clearly be seen, and
the remaining `feasible' constant-acceleration fit is also unable to
successfully reproduce this behaviour.

These residuals from the parallax model could imply that this
event is exhibiting some weak binary-source signatures (see, for example,
Alcock et al. 2001). We performed a preliminary investigation to test
this hypothesis, fitting the event with a simple binary-source
model. The basic details of this model can be found in Smith et
al. (2002b), but the version used here has two differences from the
earlier work: firstly, it
incorporates the parallax motion of the Earth (which, in most cases,
should have an effect on the light curve since the duration of this
event is well over one year\footnote{It should be noted that the
parallax effect is expected to be negligible in the case of
binary-source microlensing when the lens and sources both lie in the
bulge, so-called bulge-bulge lensing.}); and secondly, only face-on
elliptical orbits are considered. 

The best binary-source plus parallax fit that we are able to identify
has a much better $\chisq$ value ($\chisq = 162.4$) compared to all of
the previous models. However, we find a range of possible
binary-source fits with $\chisq$ less than 180. The single best-fit
solution is presented in Table \ref{tb:bin} and Fig. \ref{lc:sc33}.
This model predicts a very large value for the Einstein radius
projected into the observer plane ($\rEt = 108$ au), which implies
that the parallax motion of the Earth has little effect on the light
curve, i.e., the majority of the perturbations from the standard model
arise due to the binary nature of the source. It seems that this model
is able to fit the systematic residuals from the parallax model,
although it predicts peculiar bumps in the light curve (most notably
during the break between observing seasons). We present this model to
show that such fits are possible, but we do not attempt a more
detailed analysis of this event in this work.

If this event is indeed affected by the binary motion of the source
then a spectroscopic followup should find the
source to be a spectroscopic binary, and 
this would reduce the number of free parameters in our model.
Furthermore, if the projected separation
of the two sources can be estimated, then this would enable us to make
a direct determination of the lens mass (Han \& Gould 1997). The
source baseline magnitude (including blending)
is about $I=15.7$\,mag, and so it should 
be within easy reach of modern large telescopes.

\begin{table*}
\begin{center}
\caption{The best standard model (first row), the best parallax
model (second row), and the best constant-acceleration model (third
and fourth rows) for the OGLE-II event sc33\_4505. The parameters are
explained in \S2 \& \S3. $\theta$ is given
in units of radians. There are two sets of constant-acceleration
parameters since each $\chisq$ minimum has two degenerate fits with
differing values for $\tE$, ${\cal A}$ and $\theta$ (see Section
\ref{model} and Appendix A1).}
\label{tb:sc33}
\tiny
{
\tabcolsep=0.4\tabcolsep
\begin{tabular}{ccccccccccc}
Model & $t_0$ & $\tE$ (day) & $\u0$ & $f_0$
& $\fl$ & $\psi$ & $\rEt$ (au) & ${\cal A}$ & $\theta$ &
$\chisq$/dof \\
\hline
S &
$    647.22\pm 0.32 $ &
$    165.9^{+  8.6    }_{ -8.4    } $ &
$      0.412^{+ 0.031}_{-0.028} $ &
$   1146.3\pm 1.5 $ &
$      0.637^{+ 0.069}_{-0.058} $ &
--- & --- &
--- & --- &
490.1/179
\\ \\
P &
$    654.8^{+  3.6    }_{ -3.5    } $ &
$    191^{+  30    }_{ -21    } $ &
$     -0.213^{+ 0.054}_{-0.055} $ &
$   1144.4^{+  1.7    }_{ -2.1    } $ &
$      0.40\pm 0.11 $ &
$     3.1372^{+ 0.0086}_{-0.0104} $ &
$      6.33^{+ 0.52 }_{-0.40 } $ &
--- & --- &
217.4/177
\\ \\
A &
$    642.36\pm 0.46 $ &
$    115.38^{+ 3.44 }_{-0.88 } $ &
$      0.5596\pm 0.0014 $ &
$   1149.4^{+ 1.3}_{ -1.4 } $ &
$      1.000^{+ 0 }_{-0.064} $ &
--- & --- &
$      0.561^{+ 0.033}_{-0.024} $ &
$      4.252\pm 0.015 $ &
227.3/177
\\ \\
$\rm{A}^\prime$ &
$    642.36\pm 0.46$ &
$    166.9^{+  6.0 }_{ -2.6 } $ &
$      0.5596\pm 0.0014 $ &
$   1149.4^{+  1.3    }_{ -1.4    } $ &
$      1.000^{+ 0 }_{-0.064} $ &
--- & --- &
$      1.174^{+ 0.094}_{-0.093} $ &
$      2.268\pm 0.029 $ &
227.3/177
\end{tabular}
}
\end{center}
\end{table*}

\begin{table*}
\begin{center}
\caption{The best-fit parameters for the elliptical binary-source
model for the OGLE-II event sc33\_4505. The first seven parameters
correspond to
the usual parallax microlensing parameters (see \S3), and the final
six describe the binary nature of the source. The mass (flux) ratio,
${\cal M}$ (${\cal F}$), is defined as the mass (flux) of the first
binary source divided by the total mass (flux) of the two sources. A
detailed description of the binary-source model can be found in Smith et
al. (2002b). The errors have been omitted since they were found to be
misleading, owing to the complexity of the $\chisq$ surface.}
\label{tb:bin}
\vspace{0.3cm}
\begin{tabular}{ccccccc}
$t_0$ & $\tE$ (day) & $\u0$ & $f_0$
& $\fl$ & $\rEt$ & $\theta$\\
\hline
$ 675.2$ &
$ 140.5$ &
$ 0.573$ &
$ 1143.2$ &
$ 1.00$ &
$ 107.9$ &
$ 5.623$
\\
& & & & & &
\\
& & & & & &
\\
orbital period (day) & semi-major axis ($\rE$) & ${\cal M}$ & ${\cal F}$ & eccentricity & phase &
$\chisq/\mbox{dof}$ \\
\hline
$ 159.48$ &
$ 1.057$ &
$ 0.105$ &
$ 0.192$ &
$ 0.943$ &
$ 1.216$ &
162.415/71
\\
\end{tabular}
\end{center}
\end{table*}

\subsubsection{OGLE-1999-BUL-19 and OGLE-1999-BUL-32/MACHO-99-BUL-22}

These two events display unambiguous deviations from the standard
model. They were detected by the OGLE Early Warning System (with
OGLE-1999-BUL-32 being independently detected by the MACHO
collaboration's Alert System\footnote{
http://darkstar.astro.washington.edu/}, by which it was named
MACHO-99-BUL-22). Both of these
events also appear in the difference image analysis catalogue of 
Wo\'zniak et al. (2001), and are labelled sc40\_2895 and sc33\_3764,
respectively. For the purposes of this paper, we fit these events using
the OGLE difference image analysis data from Wo\'zniak et al. (2001).

OGLE-1999-BUL-19 (Smith et al. 2002b) displays prominent multiple peaks
that are well fit by a parallax model, while OGLE-1999-BUL-32 (Mao et
al. 2002; Bennett et al. 2002b) is the longest duration event ever
discovered. Both events are suitably fit by the parallax model, and
both show significant improvement in $\chisq$ between the standard and
parallax models.

Since the microlensing amplification for both of these events
lasts well over two years, we expect that a constant-acceleration
model may not be adequate to describe their behaviour. By fitting
with the constant-acceleration model we find that this is true;
neither of these events are well fit. The best-fit $\chisq$ values
for these events are given in Table \ref{tb:scbh}. For
OGLE-1999-BUL-19, the $\chisq$ for the constant-acceleration model
is 11420, which is far worse than the parallax model of 590.1.
Similarly, for OGLE-1999-BUL-32, the $\chisq$ for the constant
acceleration model is 464.4 compared with 278.2 for the
parallax model.

\begin{table*}
\begin{center}
\caption{The best $\chisq$ values for the two unusual events
OGLE-1999-BUL-19 (Smith et al. 2002b) and OGLE-1999-BUL-32 (Mao et
al. 2002; Bennett et al. 2002b).}
\label{tb:scbh}
\begin{tabular}{ccc}
& OGLE-1999-BUL-19 & OGLE-1999-BUL-32\\ \hline
Standard $\chisq$/dof & 17910/312 & 576.3/264\\
Parallax $\chisq$/dof & 590.1/310 & 278.2/261\\
Constant Acceleration $\chisq$/dof & 11420/310 & 464.4/261\\
\end{tabular}
\end{center}
\end{table*}

\section{Discussion}
\label{disc}

In this paper, we have illustrated the effect of acceleration
on microlensing light curves.
We studied whether the known parallax microlensing events can
be equally fitted by a model with a
constant acceleration. We found that for the shorter marginal 
parallax microlensing events, the constant acceleration is indeed
sufficient. However, for the longest events (notably,
OGLE-1999-BUL-32 and OGLE-1999-BUL-19), the constant acceleration fails
to provide a satisfactory model, so their parallax nature seems
secure.

It is important to examine whether there is an empirical way
of separating `parallax events' from `microlensing events
with acceleration'. For binary lenses it can be shown that such
constant-acceleration perturbations are unlikely to be of importance.
This can be understood by considering three regimes of binary lenses
separately: close binaries (i.e., those with separation $b \ll \rE$),
intermediate binaries ($b \sim \rE$), and wide binaries ($b \gg
\rE$). One would expect sufficiently large accelerations for close
separation binaries. However, unless the impact parameter is extremely
small (i.e. the source passes very close to the centre of mass of the
lens), it is unlikely that the resulting light curve will be
differentiable from a standard single-lens light curve (Gaudi \& Gould
1997). Even in the rare cases where such small impact parameters
occur, one would expect the primary deviations to be caused by the
binary structure, rather than the acceleration (for example, the
binary-lens event MACHO-99-BLG-47, Albrow et al. 2002). For
intermediate binaries where the separation is of the order of the
Einstein radius, one would also expect the primary deviations to arise
from the binary structure, rather than the acceleration. Even in the
absence of caustic crossings, the effect of acceleration should be
relatively small compared to the distortion of the light curve due to
the other mass (For example, typical lens geometries - with the source
lying in the bulge at a distance of approximately $8\kpc$ and the lens
lying half-way in between - predict $\rE \sim 4~{\rm au}~(M_{\rm
lens}/M_{\odot})^{1/2}$, where $M_{\rm lens}$ is the total lens mass;
therefore, for face-on orbits with separations of approximately
$1~\rE$, a typical orbital period is $P \sim 8~{\rm yr} ~ (M_{\rm 
lens}/M_{\odot})^{1/4}$, which implies that the resulting acceleration
should produce only weak perturbations to the light curve compared to
the distortion from the companion mass). For the case of wide
separation binaries, it is clear that the accelerations should be
small compared to the Earth's acceleration, since binaries with
separation much greater than $\rE$ are expected to have orbital
periods of longer than 8 years. Therefore, we conclude that the effect
of constant accelerations in binary lenses is unlikely to be of
importance.

In principle, there is a simple test to determine whether an
acceleration is due to binary motion of the source or the observer:
for the case where the acceleration is induced by the source's
binary motion, the source must show (periodic) changes in radial
velocity, which can be verified even after the event is over.
In Appendix B, we illustrate the expected radial velocities
for the simple case of circular binary orbits. We find that
the expected radial velocity amplitude is of the order of 
$35 \kms {\cal A}^{1/4} \sin i/\cos^{1/4} i $, where
$i$ is the inclination angle; there is also some
weak dependency on $\tE$, the transverse velocity of the lens, $v$,
and the masses. The expected orbital period depends on the
unknown lens transverse velocity and the inclination angle, but is
of the order of $\sim$ years. This radial velocity can, in principle, be 
measured.

\begin{figure}
\plotone{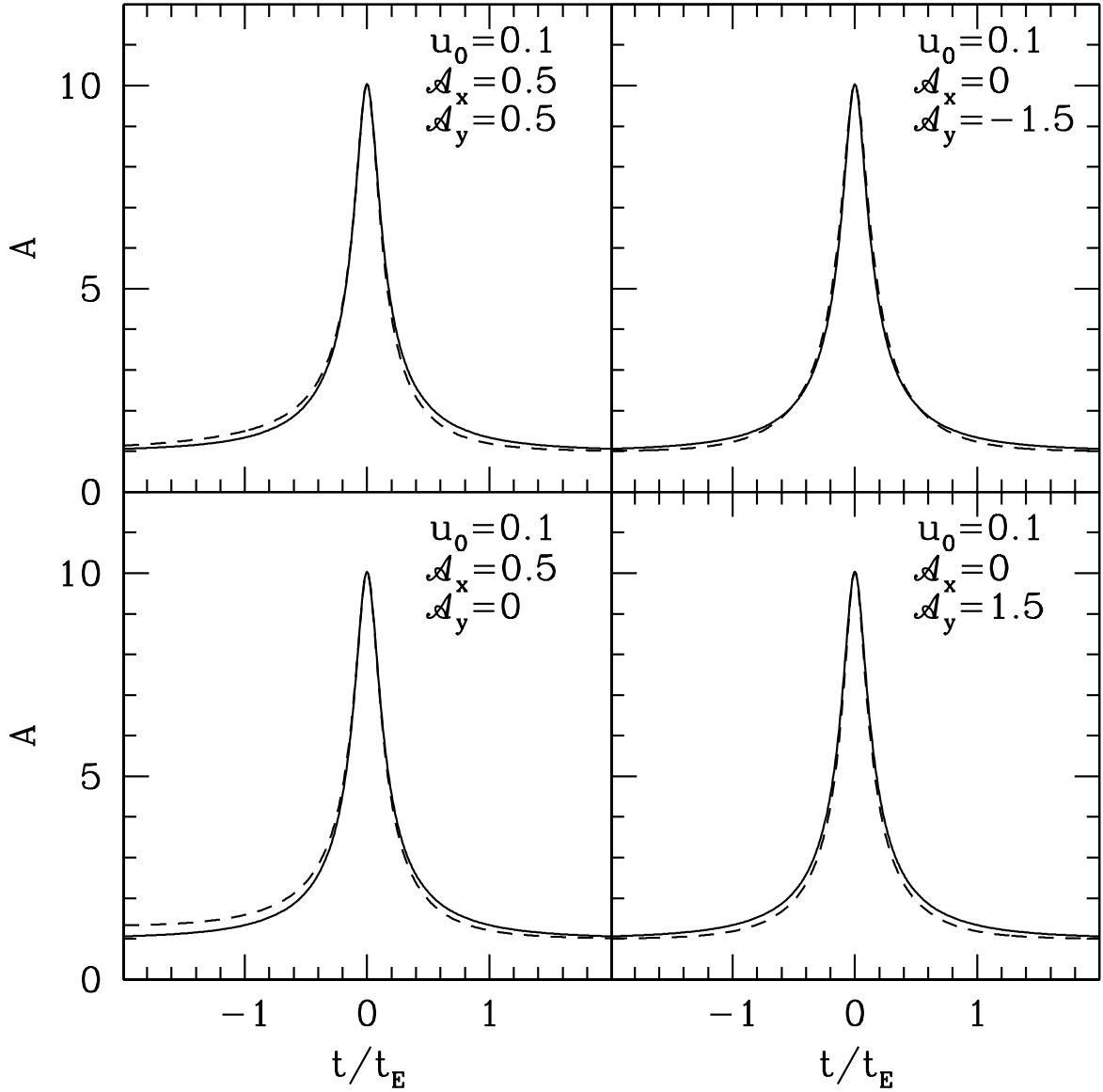} 
\caption{Four example microlensing light curves incorporating
acceleration effects. All four panels assume the same minimum impact
parameter $u_0=0.1$ in the case of no acceleration. The accelerations
along the $x$ and $y$ directions are labelled at the top right
of each panel. The source is assumed to move along the
$x$-direction in the case of no acceleration. In all panels,
the dashed and solid lines show the light curves with and
without acceleration, respectively.
}
\label{fig:model}
\end{figure}
\begin{figure}
\plotone{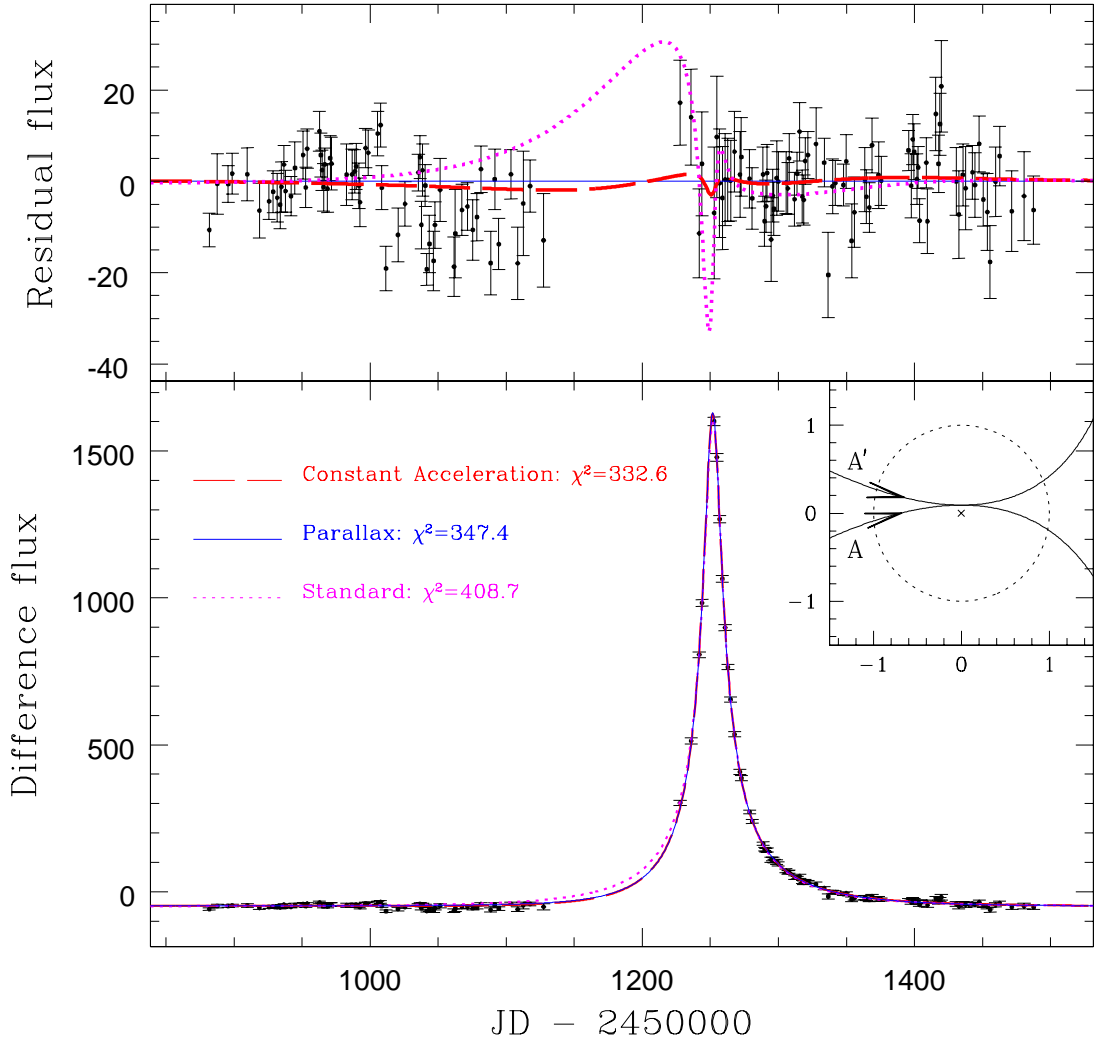} 
\caption{
The {\it I}-band light curve for OGLE-II event sc6\_2563 from
difference image analysis, with flux given in units of 10 ADU. The
data for the first season has been omitted from this plot. The
$I$-band baseline magnitude for this event is 17.7, rising to 15.1 at
the peak. The top panel shows the
residual flux (the observed data points with the parallax model
subtracted). This figure shows that the constant-acceleration model
and the parallax model produce very similar fits, with the former
model providing a slight improvement in $\chisq$. However, both
models appear to over-predict the flux around
$1000 < t < 1150$ days. The inset illustrates the two possible
trajectories for the constant-acceleration model, described by the
two fits A (bottom) and $\rm{A}^\prime$ (top) given in Table
\ref{tb:sc6}. The dotted circle represents the size of the event's
Einstein radius, $\rE$, and the axes are in units of this Einstein
radius. Both trajectories result in identical light curves,
even though the parameter values for $\tE$, ${\cal A}$ and $\theta$
differ (see Section 3 and Appendix A1 for a discussion of this
degeneracy).
}
\label{lc:sc6}
\end{figure}
\begin{figure}
\plotone{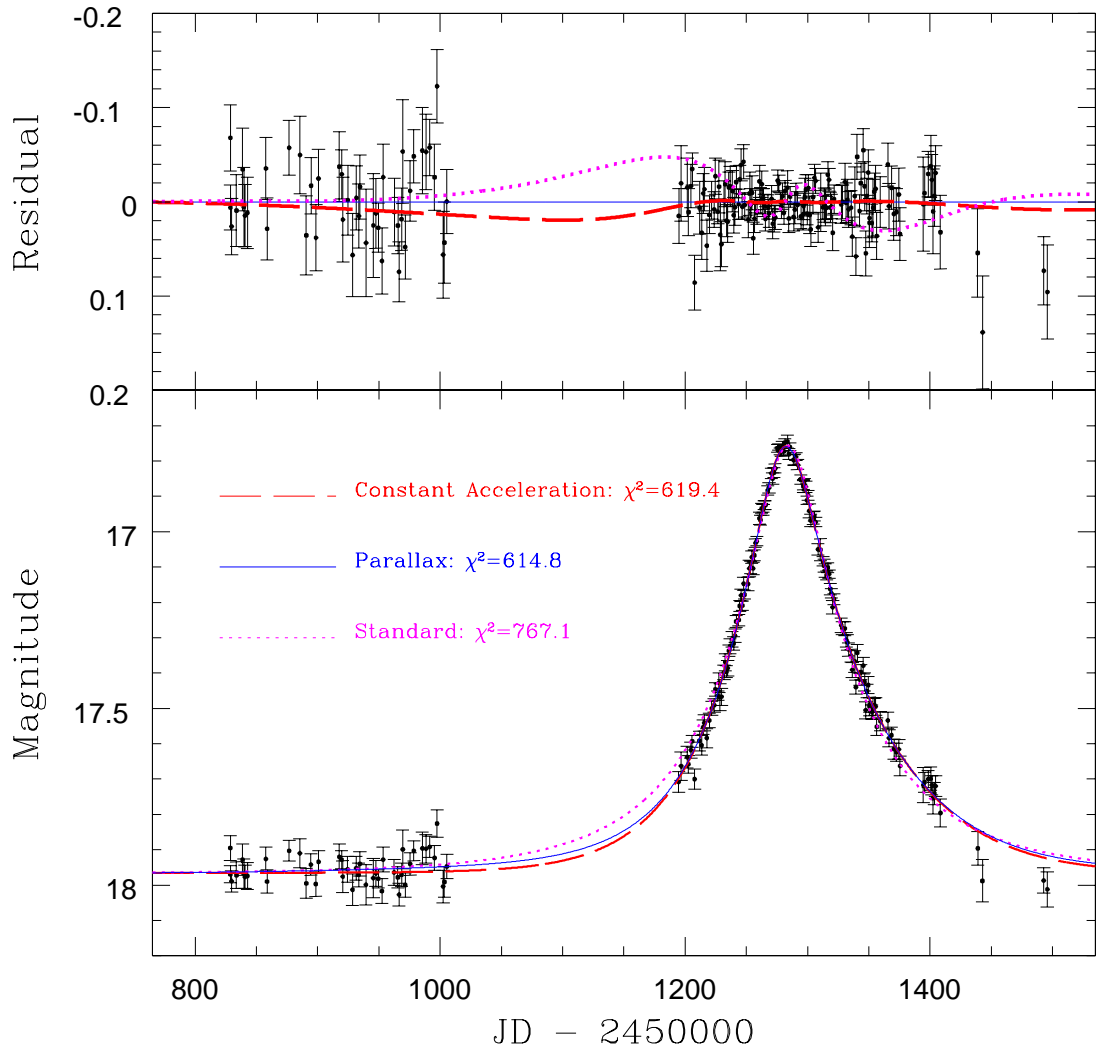}
\caption{
The {\it I}-band light curve for event OGLE-1999-CAR-1 from difference image
analysis. The data for the first season have been omitted from this
plot. The top panel shows the residual magnitude (the observed data 
points with the parallax model subtracted). This shows that the parallax
model and the constant-acceleration model produce almost identical fits.
}
\label{lc:car}
\end{figure}
\begin{figure}
\plotone{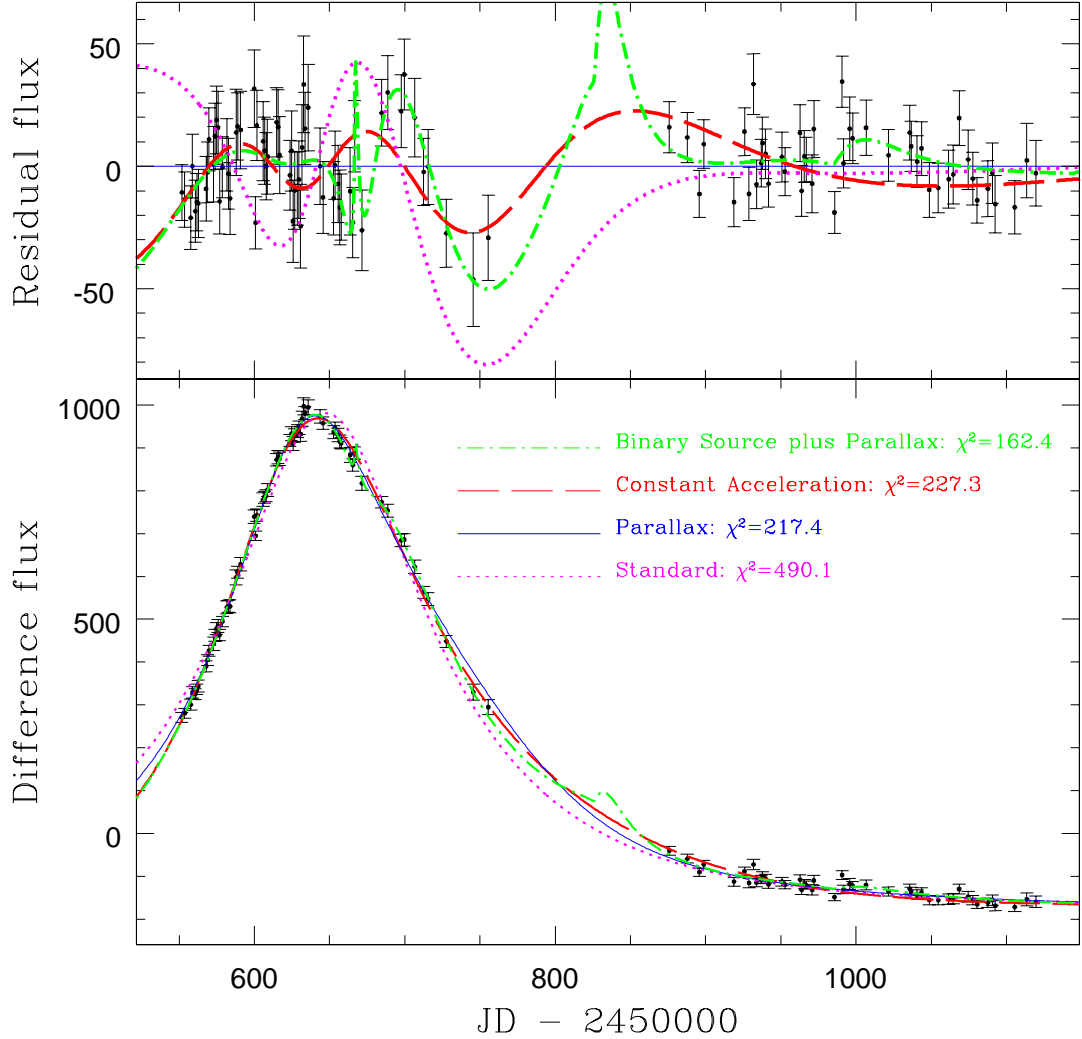}
\caption{
The {\it I}-band light curve for the OGLE-II event sc33\_4505 from
difference image analysis. The data for the third season have been
omitted from this plot. The $I$-band magnitude for this event is 15.7
at the baseline and rises to 15.0 at the peak.
The top panel shows the residual flux (the observed data points
with the parallax model subtracted). It is clear from this panel
that there are systematic residuals from the parallax fit. The constant
acceleration model appears to fit the data much better, but this still
seems to be deficient. However, the `binary-source plus parallax'
model provides the best-fit (see Section \ref{sc33}). 
Spectroscopic observations could confirm whether this event is being
affected by binary motion of the source; such observations would enable
constraints to be put on some of the model parameters, and may also
result in a direct measurement of the lens mass.
}
\label{lc:sc33}
\end{figure}

In the right two panels of Fig. \ref{fig:model}, we showed
that the light curves are symmetric when the acceleration is 
perpendicular to the source trajectory. In such cases, the
effect of acceleration may be difficult to notice. This
may even be true for some cases for which the acceleration is
not exactly perpendicular to the source motion, particularly since the
real light curves have significant gaps due to bad weather, etc.
If we force the standard fit to the microlensing light curve,
the time-scale obtained from the fit may be different from
the true value. To illustrate the effect, we generate
artificial light curves using Monte Carlo simulations. We
adopt $\tE=30$ days, and a sampling frequency of 
one observation per day; the observations lasts from $-2\tE$ to $2\tE$
centred around the peak of the light curve. The observational errors
are assumed to be 0.05 magnitudes at the baseline and scale
according to photon Poisson noise. We find that a standard fit
incorporating blending is significantly better than a fit without
blending. In other words, a microlensing light curve with constant
acceleration can mimic light curves with blending. Interestingly,
the fit with blending under-estimates the true $\tE$ by 20\% for both
acceleration values (${\cal A}=\pm 1.5$) shown in the right panels
of Fig. \ref{fig:model}. This can have important implications for the
calculation of optical depth. The optical depth, $\tau$, is estimated
from experiments using the following formula (see, for example, Han \&
Gould 1995), 
\beq
\tau = \frac{\pi}{2 N_* T_*} \sum_{i=1}^{N}
\frac{{\tE}_{,i}}{\epsilon({\tE}_{,i})},
\eeq
where $N_*$ is the total number of monitored stars, 
$T_*$ is the experiment duration in years, $\epsilon({\tE}_{,i})$ is
the detection efficiency, and ${\tE}_{,i}$ is the
Einstein radius crossing time for the $i$-th event ($i=1, \ldots, N$).
Obviously, if the time-scale is underestimated
then this will lead to an underestimation in the optical depth, and
vice versa. However, a quantitative analysis requires a detailed
knowledge about the binary parameters of the lens and the source, and
sampling frequencies in observations. This is clearly an area that
deserves further study.

\section*{Acknowledgments}

We thank Simon White for an insightful question which prompted the
study and Ian Browne for a careful reading of the paper.
We are indebted to the referee Andy Gould for a prompt and
comprehensive report.
MCS acknowledges receipt of a PPARC grant.
This project was supported by the NSF grants AST-9820314,
AST-1206213, and the NASA grant NAG5-12212 and funds 
for proposal \#09518 provided by NASA through a grant from
the Space Telescope Science Institute, which is operated by the Association of
Universities for Research in Astronomy, Inc., under NASA contract NAS5-26555.

{}

\begin{appendix}
\section{Degeneracies}
In this appendix we first describe the degeneracy that is found to
arise in the constant-acceleration model. This is followed by a
discussion of an equivalent degeneracy in the parallax model, which
can occur for short duration events where the Earth's acceleration may
be approximated as constant.

\subsection{Degeneracy in Constant-Acceleration Models}

In Section 3 we discuss the degeneracy inherent in fitting
microlensing light curves with the constant-acceleration model. It
is found that each $\chisq$ minimum has two corresponding fits, each
with differing values of time-scale, $\tE$, and the acceleration
parameters, ${\cal A}$ and $\theta$. However, the
magnitude of the physical acceleration in units of $\rE/{\rm day}^2$
(i.e., $\left| {\bf a} /\rE \right| = {\cal A}/\tE^2$) is the same for
both fits, although the direction of this vector (i.e., $\theta$)
differs. In this subsection we show analytically why this degeneracy
occurs.

Given a combination of parameters (${\tE}$, $u_0$, etc), the
magnitude of the source displacement in the lens plane (in units of
the Einstein radius) at a time $(t_0 + \delta t)$ can be determined
from equations (\ref{eq:linear}-\ref{eq:pos}),
\beq
u^2(t_0 + \delta t) = 
\left(
\hat{a} \cos{\theta} ~ \frac{(\delta t)^2}{2} + \frac{\delta t}{{\tE}}
\right)^2
+ \left(
\hat{a} \sin{\theta} ~ \frac{(\delta t)^2}{2} + u_0
\right)^2,
\eeq
where $\hat{a} = \left| {\bf a} /\rE \right| = {\cal A}/\tE^2$ is the
magnitude of the physical acceleration in units of the Einstein
radius.

This equation can be expanded to give,
\begin{eqnarray}
u^2(t_0 + \delta t) & = &
\hat{a}^2\frac{(\delta t)^4}{4}\nonumber\\
& + & \left (
\hat{a} \frac{\cos{\theta}}{{\tE}}
\right) (\delta t)^3\nonumber\\
& + & \left (
\frac{1}{{\tE}^2} + u_0 \hat{a} \sin{\theta}
\right) (\delta t)^2\nonumber\\
& + &
u_0^2.
\label{Aeq:pos}
\end{eqnarray}
For two degenerate solutions $i$ and $j$,
the magnitude of the source displacement, $u_i(t_0+\delta t)$ and
$u_j(t_0+\delta t)$, must be the same for all values of $\delta t$
(since this is required to produce identical light curves), i.e.,
$u_i^2(t_0 + \delta t) = u_j^2(t_0 + \delta t)$ for all $\delta
t$. Since this is true for all values of $\delta t$, using equation
(\ref{Aeq:pos}) we can equate powers of $\delta t$. This gives the
following four equations,
\begin{eqnarray}
\hat{a}_i&=& \hat{a}_j
\label{Aeq:1}\\
\hat{a}_i \frac{\cos{\theta}_i}{{\tE}_{,i}}
&=&
\hat{a}_j \frac{\cos{\theta}_j}{{\tE}_{,j}}
\label{Aeq:2}\\
\frac{1}{{\tE}_{,i}^2} + u_{0,i} \hat{a}_i \sin{\theta}_i
&=&
\frac{1}{{\tE}_{,j}^2} + u_{0,j} \hat{a}_j \sin{\theta}_j
\label{Aeq:3}\\
u_{0,i}^2
&=&
u_{0,j}^2.\label{Aeq:4}
\end{eqnarray}
From equations (\ref{Aeq:1}) and (\ref{Aeq:4}) we can see that both
$\hat{a}$ and $u_0$ must be constant for degenerate solutions
(the sign of $u_0$ is irrelevant due to the symmetry of the
acceleration model; by taking only $u_0 > 0$ one still obtains the
full set of solutions).
Also, from equations (\ref{Aeq:2})
and (\ref{Aeq:3}) we find that the quantities $[\cos\theta / \tE]$ and
$[1/( \tE )^2 + u_0 \hat{a} \sin \theta]$ must also be constant for
degenerate solutions. Therefore, the following two equations must hold
for {\it every} degenerate solution,
\begin{eqnarray}
\frac{\cos\theta}{\tE} = c_1,\\
\frac{1}{\tE^2} + u_0~\hat{a}~\sin\theta = c_2,
\end{eqnarray}
where $c_1$ and $c_2$ are constants defined for any given $\chisq$
minimum. The parameter $\theta$ can then be eliminated from these two
equations to give the following expression for $\tE$,
\beq \label{Aeq:tE}
f(w) =
(\hat{a} u_0 c_1)^2 {w}^3 + (c_2^2 - \hat{a}^2u^2_0) w^2 - 2c_2 w + 1
= 0,
\eeq
where $w \equiv \tE^2$.
We see that the above equation is a cubic equation
in $w$. As the coefficients are all real, the number of
real solutions is either one or three. Recall that since $w=\tE^2$,
physical solutions to equation (\ref{Aeq:tE}) must clearly correspond
to positive values of $w$. 

As $f(-\infty)=-\infty$, and $f(w=0)=1$, we must have
one negative real solution for $w$, which is unphysical and should be
discarded\footnote{
It is possible to have three negative solutions to this equation,
although this possibility can be ignored since this would result in no
positive (i.e. physical) solutions
}. It is easy to show that the number of positive (i.e. physical)
solutions for 
$w$ must be either zero or two\footnote{Although in rare cases, the
two positive solutions can be identical}. This follows because
$f(w=0)=1$ and $f(w=+\infty)=+\infty$. If $f(w)$ is everywhere
positive for $w>0$, then the number of positive solutions is obviously
zero. If, however, $f(w)$ is negative at some point ${w}_{1} (>0)$,
then there must be one positive solution in the region 0 to ${w}_{1}$,
and another positive solution in the region 
${w}_{1}$ to $+\infty$. This therefore immediately implies that 
if there is one good constant-acceleration fit (i.e. one physical
solution for $w$), then there must be another degenerate fit with the
same $\chi^2$, just as we have found numerically. These two fits have
identical values of $u_0$ and $\hat{a}={\cal A}/\tE^2$, but different
values of ${\cal A}$, $\theta$ and $\tE$.

\subsection{A new parallax degeneracy}

A consequence of this constant-acceleration degeneracy is that a
similar degeneracy should arise for short-duration events that exhibit
weak parallax signatures. This is because the acceleration in the
parallax model can be approximated as constant provided that the
duration of the event is sufficiently short. In the following
subsection we consider this new degeneracy and provide an example
using the event sc6\_2563, which is introduced in Section \ref{sc6}.

There are currently well-known degeneracies inherent in parallax
microlensing: for example, Gould, Miralda-Escude \& Bahcall (1994)
discussed the continuous degeneracy that occurs for events with
extremely weak parallax signatures where only one component of the
relative lens velocity, $ {\bf \tilde{v}} $, is measurable; another
four-fold discrete degeneracy was identified in the case of satellite
parallax measurements (Refsdal 1966; Gould 1994), although it was
shown that this problem can be resolved using higher-order effects
(Gould 1995).

As we have discussed above, in the case of the weak candidate parallax
events, such as sc6\_2563 (see section \ref{sc6}), the best
constant-acceleration fit mimics the best parallax
fit. We also found that each $\chisq$ minima for the
constant-acceleration model possesses two degenerate fits, each with
identical values for impact parameter, $u_0$ and acceleration
magnitude, $\hat{a}={\cal A}/\tE^2$, but different values of ${\cal
A}$, $\theta$ and $\tE$. However, since the above
constant-acceleration formalism is rotationally invariant, it is
possible to rotate one of these degenerate constant-acceleration fits
so that both degenerate fits have their acceleration vectors
co-aligned. Therefore, if one of these best-fit constant-acceleration
solutions mimics the existing parallax fit, then the degenerate
counterpart (once rotated through the required angle so that the
acceleration is also in the same direction as the Earth's acceleration)
should likewise mimic an equivalent parallax fit. Since this new parallax
fit must be different from the existing one, this implies that there
should be two degenerate parallax fits. Obviously, this argument
only holds in the regime of constant-acceleration, i.e., this would
not apply to parallax events where the duration is sufficiently long
to invalidate the constant-acceleration approximation.

We illustrate this new degeneracy by providing an example using the
event sc6\_2563, which was introduced in Section \ref{sc6}. To
identify any degenerate solutions in the parallax parameter space we
fit this event 1000 times, taking random values for the initial
parameter guesses. From this we identify another solution in the
parameter space that has a slightly worse $\chisq$ value than the
previously identified best-fit parallax solution. Both fits are
presented in Table \ref{tb:sc6}. If the $\chisq$ values are
renormalised to enforce the $\chisq$ per degree of freedom to be unity
for the best-fit parallax model, then the difference in $\chisq$ is
less than 0.5. Unlike the two degenerate constant-acceleration
fits, these parallax fits have different values for the magnitude of
the impact parameter, $u_0$. However, this is due to the difference in
the definition of $u_0$ in the two models.

\begin{figure}
\plotone{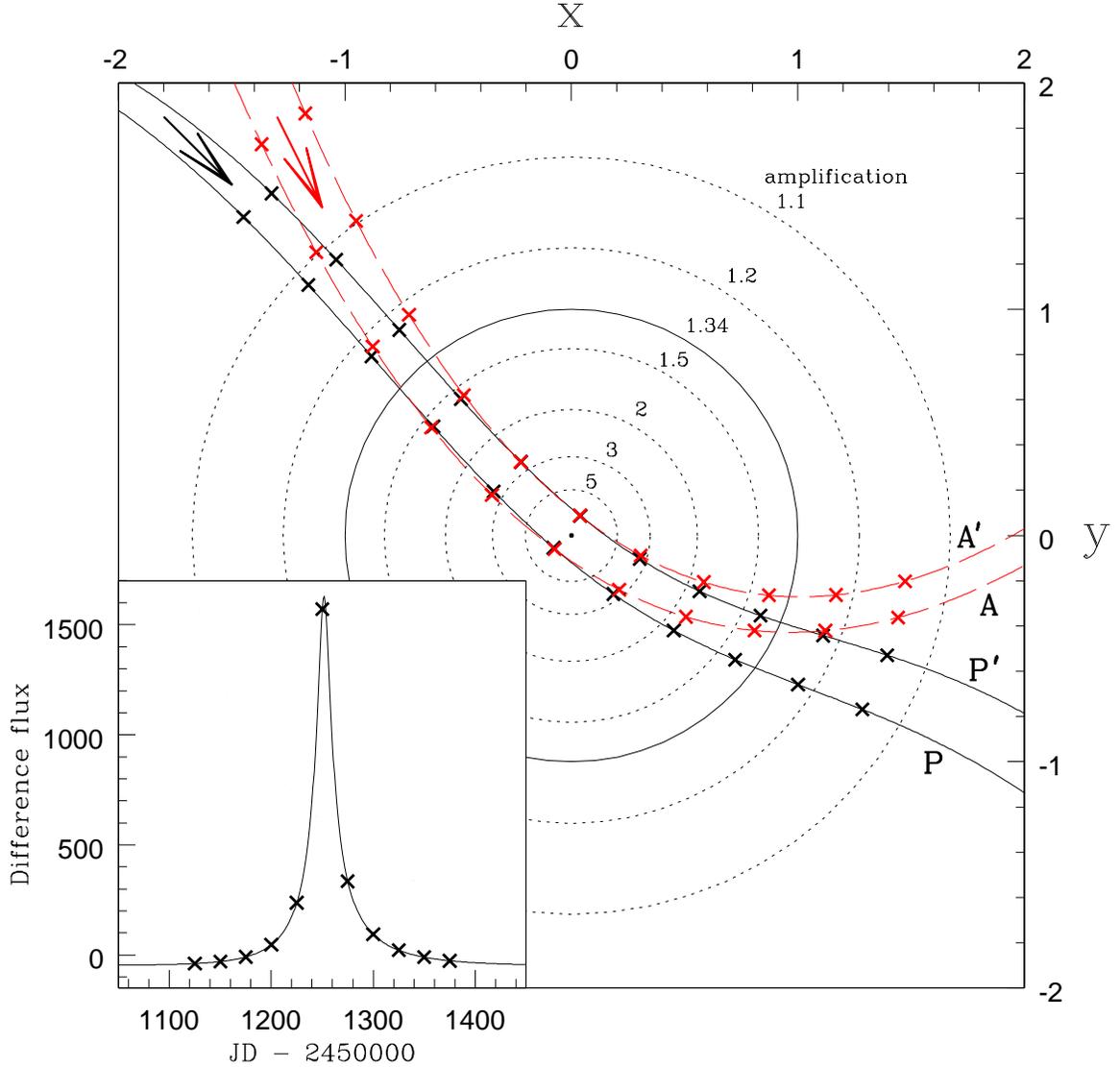}
\caption{
The degenerate parallax (solid lines) and constant-acceleration
(dashed lines) trajectories for the OGLE-II event sc6\_2563. These
trajectories are described in the lens plane and correspond to the
paths taken by the lens relative to the observer-source
line-of-sight (i.e. the origin). Each degenerate fit is labelled
according to Table \ref{tb:sc6}. 
Note that the $A^\prime$ trajectory has been rotated so that its
acceleration is pointing in the same direction as the $A$ trajectory
(the rotation does not affect the light curve).
The contours of constant
amplification are represented by the dotted circles (since the
amplification is dependent on the separation of the lens from the
observer-source line-of-sight; see eq. \ref{eq:mag}), and the 
Einstein radius is given by the solid circle. The inset shows the
light curve for this event. The crosses on each trajectory denote 25
day intervals, as labelled on the inset light curve. Note that all of the
trajectories presented here result in (practically) the same light
curve.
}
\label{fig:traj}
\end{figure}

Figure \ref{fig:traj} shows the trajectories in the lens plane for the
two parallax fits, along with the two degenerate constant-acceleration
fits. Note that the $A^\prime$ trajectory has been rotated so that its
acceleration is pointing in the same direction as the $A$ trajectory
(the rotation does not affect the light curve).
Although each parallax trajectory deviates from its corresponding
constant-acceleration counterpart, it can be seen that the distance
from the origin (i.e., the quantity that determines the magnification;
the parameter $u$ in eq. \ref{eq:mag}) is very similar. As expected, this
figure shows that the difference between the constant-acceleration and
parallax trajectories only becomes apparent away from the point of
closest approach, i.e. when the direction of the Earth's acceleration
has changed. This corresponds to the wings of the light curve, but
since the magnification is only weak in this region the resulting
deviations in the light curve are difficult to detect.

We also investigated the event sc33\_4505, which is introduced in
Section \ref{sc33}. This event exhibits more prominent deviations, 
and the constant-acceleration model is found to provide a
significantly worse fit than the parallax model (see Table
\ref{tb:sc33}). Again, we fit this event 1000 times to try and identify
any degeneracy that might occur for the parallax model, but we are
unable to find any such degenerate fits. The next-best parallax fit has
$\chisq=227.9$, which is more than $3\,\sigma$ away from the best-fit
value of $\chisq=217.4$. Therefore, we conclude that there appears to
be no such parallax degeneracy for this event, owing to the fact that
the deviations in the light curve are relatively prominent, i.e.,
these deviations are unable to be reproduced by a
constant-acceleration fit.

\section{Radial Velocity}

Here we estimate the expected radial velocity for the case where
the acceleration is induced by the binary motion of
the source. For simplicity, we
consider the case where both orbits are circular. Let
us denote the mass of the lensed source star as $M$ and
that of its binary companion as $\Mp$, the separation between
the stars as $d$, and the inclination of the orbital plane as $i$
($i=90^\circ$ implies an edge-on orbit).

In the centre-of-mass rest frame, the maximum radial velocity is given by
\begin{equation} \label{eq:A1}
v_r = \left({G {\Mp}^2 \over d} {1 \over M+\Mp}\right)^{1/2} \sin i.
\end{equation}
The physical acceleration of the lensed source due to its companion
is given by ${G\Mp/d^2}$. The component perpendicular to the
line-of-sight is $a=(G\Mp/d^2) (1-\sin^2 \phi \sin^2 i)^{1/2}$; here
$\phi$ is the angle of the position of the lensed star from the
line where the orbital plane and the plane of sky intersect. 
Notice that our constant acceleration assumption requires that $\phi$
does not change significantly during the lensing event, i.e. for
circular orbits $\Delta\phi << 1$ implies that the period of the binary
$P >> 2\pi\tE$.
Combining the expression of $a$ with the definition
of ${\cal A}$, we have
\begin{equation} \label{eq:A2}
{\cal A} \equiv {a \over \rE/\tE^2} = {G\Mp \over d^2} {\tE^2 \over \rE}
(1- \sin^2\phi \sin^2 i)^{1/2}.
\end{equation}
Cancelling out $d$ from eqs. (\ref{eq:A1}) and
(\ref{eq:A2}), we obtain the maximum radial velocity
\begin{equation}
v_r = \left({\Mp G v \over \tE}\right)^{1/4}
{\cal A}^{1/4} \left({\Mp \over M+\Mp}\right)^{1/2} {\sin i \over
\left(1- \sin^2\phi \sin^2 i\right)^{1/8}},~
\end{equation}
where $v(=\rE/\tE)$ is the lens transverse velocity perpendicular to the
observer-source line. Numerically, we have
\begin{equation}
v_r = 35 \kms
\left({\Mp \over M_\odot} ~~
{ 100 {\rm days} \over \tE} ~~
{v \over 100\kms}\right)^{1/4} 
{\cal A}^{1/4} \left({\Mp \over M+\Mp}\right)^{1/2} {\sin i \over
\left(1- \sin^2\phi \sin^2 i\right)^{1/8}}.
\end{equation}
Notice that the dependence on the masses, ${\cal A}$ and
the lens transverse velocity, $v$, are fairly weak. 

From Kepler's third-law, the period of the binary 
is given by $P=2\pi d^{3/2} G^{-1/2}(M+\Mp)^{-1/2}$. Substituting
$d$ from eq. (\ref{eq:A2}) into $P$, we obtain
\begin{equation}
P = 221 \,{\rm days}
\left({\Mp \over M_\odot}\right)^{1/4} ~~
\left({ {\tE \over 100\,{\rm days}}~~
{100\kms \over v}}\right)^{3/4}
{\cal A}^{-3/4} \left({\Mp \over M+\Mp}\right)^{1/2}
\left(1- \sin^2\phi \sin^2 i\right)^{3/8}.
\end{equation}
The dependence of period on the unknown lens transverse velocity
and the inclination is fairly sensitive. For events that show constant
accelerations, the transverse velocity may be fairly low and hence
the period can be of the order of years.

\end{appendix}

\clearpage

\end{document}